\documentclass[aps,pra,reprint,superscriptaddress,showpacs]{revtex4-1}

\usepackage{graphicx}% Include figure files
\usepackage{dcolumn}% Align table columns on decimal point
\usepackage{bm}% bold math
\usepackage{amsmath,amssymb}
\usepackage{physics}
\DeclareMathOperator{\sinc}{sinc}
\usepackage{mathrsfs}
\usepackage{mhchem}
\usepackage{siunitx}
\usepackage{color}

\begin{document}

%\preprint{APS/123-QED}

\title{Seeded and unseeded high order parametric down conversion}% Force line breaks with \\

\author{Cameron Okoth}
\email{cameron.okoth@mpl.mpg.de}
\affiliation{Max Planck Institute for the Science of Light, Staudtstra{\ss}e 2, 91058 Erlangen, Germany}
\affiliation{University of Erlangen-N\"urnberg, Staudtstra{\ss}e 7/B2, 91058 Erlangen, Germany}
\author{Andrea Cavanna}
\affiliation{Max Planck Institute for the Science of Light, Staudtstra{\ss}e 2, 91058 Erlangen, Germany}
\affiliation{University of Erlangen-N\"urnberg, Staudtstra{\ss}e 7/B2, 91058 Erlangen, Germany}
\author{Nicolas Joly}
\affiliation{University of Erlangen-N\"urnberg, Staudtstra{\ss}e 7/B2, 91058 Erlangen, Germany}
\affiliation{Max Planck Institute for the Science of Light, Staudtstra{\ss}e 2, 91058 Erlangen, Germany}
\author{Maria Chekhova}
\affiliation{Max Planck Institute for the Science of Light, Staudtstra{\ss}e 2, 91058 Erlangen, Germany}
\affiliation{University of Erlangen-N\"urnberg, Staudtstra{\ss}e 7/B2, 91058 Erlangen, Germany}
\affiliation{Department of Physics, M. V. Lomonosov Moscow State University, Leninskie Gory, 119991 Moscow, Russia} %Lines break automatically or can be forced with \\

\date{\today}% It is always \today, today,

\begin{abstract}

Spontaneous parametric down conversion (SPDC) has been one of the foremost tools in quantum optics for over five decades. Over that time it has been used to demonstrate some of the curious features that arise from quantum mechanics. Despite the success of SPDC, its higher-order analogues have never been observed, even though it has been suggested that they generate far more unique and exotic states than SPDC. An example of this is the emergence of non-Gaussian states without the need for post-selection.
%The first experimental observation of spontaneous parametric down conversion (SPDC) happened over five decades ago. With these initial experiments the field of non-linear quantum optics was established. Despite the many successes of second-order SPDC, it remains the only SPDC process to have been experimentally observed. Although SPDC is one of the main tools in quantum optics, it has been suggested that its high-order analogues should generate states with far more exotic quantum features. A good example of this is the emergence of non-Gaussian Wigner functions for high-order states.
Here we calculate the expected rate of emission for $n^{th}$-order SPDC with and without external stimulation (seeding). 
Focusing primarily on third-order parametric down-conversion (TOPDC), we estimate the photon detection rates in a rutile crystal, for both the unseeded and seeded regimes.
%We  find that the seed introduces an enhancement factor of roughly $10^5$ to the rate of emission when compared to the spontaneous case.   

\end{abstract}

\maketitle

%\tableofcontents

\section{\label{sec:intro}Introduction}

Nonlinear optical effects are so far the most convenient tool for generating nonclassical states of light. For instance, parametric down-conversion \cite{harris1967observation,magde1967study,akhmanov1967quantum} and four-wave mixing \cite{slusher1985observation} are widely used for producing photon pairs, single photons (through heralding \cite{hong1986experimental,rarity1987observation}), quadrature squeezed light \cite{wu1986generation} and twin beams \cite{heidmann1987observation}. Meanwhile, there are attempts to realize higher order nonlinear effects, leading to the generation of photon triplets (or third-order squeezing \cite{banaszek1997quantum,elyutin1990three,felbinger1998oscillation}). Despite a large number of proposals and theoretical papers \cite{corona2011experimental,gonzalez2018continuous,bencheikh2007triple,moebius2016efficient,akbari2016third,cavanna2016hybrid}, the direct decay of a pump photon into three daughter photons, further called third-order parametric down-conversion (TOPDC), has not yet been realized experimentally.

Photon triplet states have indeed been obtained in experiment using cascaded quadratic nonlinear effects \cite{hubel2010direct} or accidentally overlapping photon pairs emitted through parametric down-conversion \cite{mosley2008heralded}. However, the statistics of light emitted through these effects is very different from the statistics of photons being generated by TOPDC: for instance, in the `cascaded' experimental realisations, there is a strong asymmetry between the photon numbers in the three output beams.
%The resulting photon statistics differs from the one expected for TOPDC. 
Recently, rather high rates of photon triplet generation have been reported by using exciton and biexciton transitions in coupled quantum dots \cite{khoshnegar2017solid} but it is not clear to what extent this process is similar to TOPDC.

Difficulties arise when generating triplet photons via TOPDC or any high order effect due to the fundamentally low efficency of such processes. A natural step towards the observation of TOPDC is to seed (stimulate) the emission of one of the photons in the three photon state. An important breakthrough in this direction has been made in \cite{douady2004experimental}, where two seeding beams were used. However, no nontrivial photon statistics could be observed at the output in this case.

In this work we give a general description of  $n^{th}$-order SPDC and compare the efficencies of lower-order processes and higher-order processes. We then describe how the seeding of an $n^{th}$-order process using a strong coherent source affects the rate of photon emission and changes the fundamental properties of the emitted radiation. In particular, we show that seeded TOPDC generates a two-photon state and not a three-photon state like spontaneous TOPDC. Despite the loss of the three-photon state, we still believe stimulated TOPDC is an interesting effect to observe as it can be used as a way to study spontaneous TOPDC, the same way as stimulated emission tomography \cite{liscidini2013stimulated} is used to characterize the properties of SPDC.

The paper is structured as follows. In Section~\ref{sec:rateofn} we derive the rate of $n$-photon SPDC starting from Fermi's golden rule.
In Section~\ref{sec:phasematching} we analyse the phase matching function and distinguish between two regimes: broadband detection and narrowband detection. In Section~\ref{sec:diffcount} we relate the emission rates of high order processes to the emission rates of lower order processes. The effect of seeding is considered in Section~\ref{sec:seed}, and the spectral properties of seeded and unseeded TOPDC emission are explored in Section~\ref{sec:unseededcalc} and Section~\ref{sec:seededcalc}. In Section~\ref{sec:estimates} we present an estimate for the expected triplet, double and single photon count rates for seeded and unseeded TOPDC in  rutile. We conclude (Section~\ref{sec:conc}) with a discussion of the main results. 

\section{\label{sec:rateofn} The rate of an $n$ photon transition per mode}

An $n^{\text{th}}$-order parametric down conversion process involves the transition of a single pump photon to a state of $n$ photons occupying, in the general case, $n$ modes. Using the approach outlined in \cite{klyshko1988photons}, we calculate the rate of an $n$ photon transition per mode using Fermi's golden rule,
\begin{equation}\label{eq:Fermi}
    \Gamma^{(n)}=\frac{2\pi}{\hbar^2}\abs{ \bra{\bra{1}}_{n}\hat{H}^{(n)}{\ket{\ket{0}}_{n}}}^{2}\delta (\Delta \omega^{(n)}),
\end{equation}
where
\begin{equation}\label{eq:energycons}
    \Delta \omega^{(n)}=\omega_p-\sum_{ i=1}^{n}\omega_i,
\end{equation}
the subscript $i$ denotes the mode with frequency $\omega_i$ and wavevector $\vec{k}_i$, the subscript $p$ denotes the pump mode and $\ket{\ket{0}}_{n}$,$\ket{\ket{1}}_{n}$ signify the $n$ dimensional vacuum state and the product state of $n$ modes each populated by a single photon, respectively. $\hat{H}^{(n)}$ is the Hamiltonian of an $n^{\text{th}}$-order nonlinear perturbation, which in an isotropic medium can be described macroscopically as
\begin{equation}\label{eq:Hamiltonian}
     \hat{H}^{(n)}=-\epsilon_0 \frac{n!\chi^{(n)} }{2^{n}} \int F(\vec{r}) E_p^{(+)}(\vec{r})\prod_{i=1}^{n} E_i^{(-)}(\vec{r}) d^{3}\vec{r}+ h.c. ,
\end{equation}
where $\epsilon_0$ is the vacuum permittivity, $\chi^{(n)}(\vec{r})$ is the $n^{th}$-order susceptibility, which has been separated into the effective susceptibility $\chi^{(n)}$ and its spatial distribution $F(\vec{r})$, a function that is dimensionless and takes a maximum value of unity. $E_i^{(+/-)}(\vec{r})$ is the positive/negative frequency electric field component of the $i^{th}$ mode. The pump field $E_p^{(+)}(\vec{r})$, using the correspondence principle, can be described classically in the limit of large photon numbers. Assuming that the pump propagates in the $z$ direction, the classical and quantised electric fields are 
\begin{align}\label{eq:electricfield}
    E_{p}^{(+)}(\vec{r})=&A_p(x,y)\sqrt{\frac{2 I_p}{\epsilon_0 c n_p}}e^{\imath k_p\cdot z}, \nonumber \\
    E_{i}^{(-)}(\vec{r})=& \sqrt{c_i} a_i^{\dagger} e^{-\imath \vec{k}_i \cdot \vec{r}},
		\end{align}
respectively. Here, 
\begin{equation}\label{eq:electricfieldamp}
    c_i=- \frac{\hbar \omega_i v_i}{2 V_q \epsilon_0 c n_i},
\end{equation}
$I_p$ is the pump intensity, $c$ is the speed of light in the vacuum, $a_i^{\dagger}$ is the creation operator of mode $i$, $V_q$ is the quantisation volume, $n_i$ and $v_i$ are the refractive index and group velocity of mode $i$, respectively. $A_p(x,y)$ is the transverse spatial distribution of the pump which is dimensionless and has a maximum value of unity. The temporal part of the electric fields is accounted for in Fermi's golden rule; therefore, only the spatial part of the fields is considered. 
Combining Eqs.~(\ref{eq:Hamiltonian}) and (\ref{eq:electricfield}) we obtain
\begin{equation}
     \hat{H}^{(n)}=\gamma^{(n)}f\left(\Delta k^{(n)}\right)\prod_i^n \sqrt{c_i} a_i^{\dagger} + h.c.,
\end{equation}
where
\begin{equation}\label{eq:gamma}
    \gamma^{(n)}=-\frac{n!\chi^{(n)}}{2^n}\sqrt{\frac{2 I_p \epsilon_0}{c n_p}},
\end{equation}
$f\left(\Delta k^{(n)}\right)$, which we shall call the phase matching function, is given by
\begin{equation}\label{eq:phasematchingfunc}
    f\left(\Delta k^{(n)}\right)=\int F(\vec{r})A_p(x,y)e^{\imath \Delta \vec{k}^{(n)} \cdot \vec{r}} d^3 \vec{r},
\end{equation}
and 
\begin{equation}\label{eq:phasemismatch}
    \Delta \vec{k}^{(n)}= \vec{k}_p-\sum_i^n \vec{k}_i
\end{equation}
is the wavevector mismatch. The phase matching function is of importance as it couples all modes, involved in the interaction, together. From Eq.~(\ref{eq:Fermi}) the rate of transition from the vacuum state to an $n$ photon state is 
\begin{equation}\label{eq:transpermode}
     \Gamma^{(n)}=\frac{2\pi}{\hbar^2}\left[\gamma^{(n)}\right]^2 D(\Delta \vec{k}^{(n)}, \Delta \omega^{(n)}) \prod_{ i=1}^n \abs{c_i},
\end{equation}
where
\begin{equation}\label{eq:probabilitydist}
    D(\Delta \vec{k}^{(n)}, \Delta \omega^{(n)})=\abs{f\left(\Delta k^{(n)}\right)}^2 \delta( \Delta \omega^{(n)}). 
\end{equation}
Eq.(\ref{eq:transpermode}) gives the rate of transition into a single set of $n$ modes. The total transition rate is given by the number of transitions in the interval between $\vec{k}_i$ and $\vec{k}_i+d\vec{k}_i$. In three dimensional wavevector space each state occupies a $k$-space volume of $\dfrac{(2\pi)^{3}}{V_q}$. Therefore the transition rate to an 
$n^{\text{th}}$-order state in the intervals $d\vec{k}_i$ is 
\begin{equation}\label{eq:totalcountratecon}
    dN^{(n)}=\Gamma^{(n)} \frac{V_q^n}{(2\pi)^{3n}}\prod_{ i=1}^n d\vec{k}_i,
\end{equation}
which gives
\begin{equation}\label{eq:totalcountrateexp}
    dN^{(n)}=\frac{2\pi}{\hbar^2}\left[\gamma^{(n)}\right]^2 D(\Delta \vec{k}^{(n)}, \Delta \omega^{(n)}) \frac{V_q^n}{(2\pi)^{3n}} \prod_{i=1}^n \abs{c_i} d\vec{k}_i.
\end{equation}

\section{\label{sec:phasematching}Phase Matching Function}

The phase matching function and the energy conservation form a distribution $D(\Delta \vec{k}^{(n)}, \Delta \omega^{(n)})$ that limits the number of available final states in which the initial state can transition. The shape of the distribution dictates the spectral properties and the degree of entanglement of the generated photons. The energy conservation term, due to the near instantaneous response of $n^{\text{th}}$-order SPDC, is given by a delta function following from Fermi's golden rule. The phase matching term, given by Eq.~(\ref{eq:phasematchingfunc}), can be normalised by defining the interaction volume as $V=\int F(\vec{r}) A_p(x,y) d\vec{r}$. Then,
\begin{equation}\label{eq:normalisedphasematchingfunc}
     \abs{f\left(\Delta \vec{k}^{(n)}\right)}^2=\abs{\mathscr{F}\left(F(\vec{r})A_p(x,y)\right)}^2\equiv  \abs{\widetilde{f}\left(\Delta \vec{k}^{(n)}\right)}^2 V,
\end{equation}
where $\mathscr{F}$ denotes the Fourier transform. The function $|\widetilde{f}\left(\Delta k^{(n)}\right)|^2$ is constant and dimensionless when integrated over all  $k$-space.

\begin{figure}[htbp]
  \centering
  \includegraphics[width=8cm]{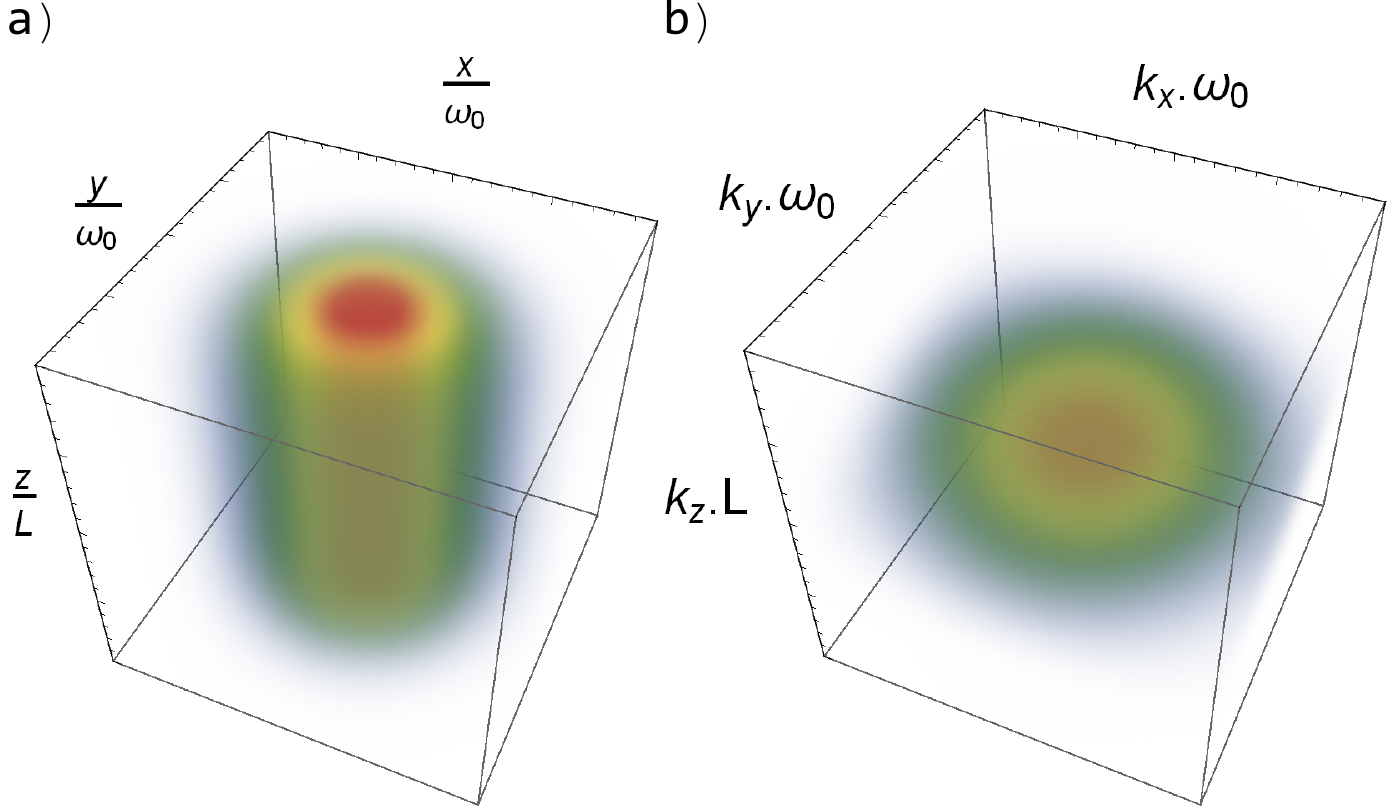}
\caption{The distributions of (a) the interaction volume in which the non-linear process takes place and (b) the phase matching function in the reciprocal ($k$) space.}
\label{fig:ftgaussian}
\end{figure}

We assume the pump to be a Gaussian beam with the waist $w_0$,
\begin{equation}\label{eq:pumpbeam}
    A_p(x,y)=\exp\left(-\left[\frac{x^2+y^2}{w_0^2}\right]\right),
\end{equation}
and a Rayleigh length much larger than the length $L$ of the nonlinear medium. If the spatial distribution of the susceptibility is uniform throughout the medium, then 
\begin{equation}\label{eq:susceptibilitywidth}
    F(\vec{r})= \Pi \left(\frac{z}{L}\right),
\end{equation}
where $\Pi$ is a rectangular function. Such a distribution is shown in Fig.~\ref{fig:ftgaussian}(a). From Eqs.~(\ref{eq:normalisedphasematchingfunc}), (\ref{eq:pumpbeam}) and (\ref{eq:susceptibilitywidth}) the squared modulus of the normalized phase matching function is 
\begin{align}\label{eq:phasematchingGaussianbeam}
    \abs{\tilde{f}\left(\Delta \vec{k}^{(n)}\right)}^2=& V\exp \left(-\frac{\Delta k_x^2+\Delta k_y^2}{4}w_0^2\right)\times \nonumber \\
    &\sinc^2\left(\frac{\Delta k_z L}{2}\right),
\end{align}
which is shown in Fig.~\ref{fig:ftgaussian}(b).

As mentioned before, when integrated over all $k$-space this function is a dimensionless constant. For this reason, whenever a convolution with a broader function is considered, we will replace Eq.~(\ref{eq:phasematchingGaussianbeam}) with
\begin{equation}\label{eq:pmdelta}
     \abs{\tilde{f}\left(\Delta \vec{k}^{(n)}\right)}^2\rightarrow(2\pi)^3  \delta^{(3)}(\Delta \vec{k}^{(n)}).
\end{equation}
Further on, we will distinguish between two detection regimes. The first is when the detection bandwidth $\text{D}_i$ is broader than the phase matching function $V^{-1}$. This we will refer to as the broadband regime. The second is when the detection bandwidth is narrower than the phase matching function. This we will call the narrowband regime. In the broadband regime, without loss of generality, we will use Eq.~(\ref{eq:pmdelta}) to represent the phase matching function and in the narrowband case we will use Eq.~(\ref{eq:phasematchingGaussianbeam}). 

\section{\label{sec:diffcount}Comparison of high order processes to low order processes}

Unlike second-order SPDC, where the final states are well defined by energy and momentum conservation, the final states of higher-order SPDC are almost continuous in $k$-space as the number of ways to fulfill the phase matching condition increases with the process order $n$. In this situation the number of states that can be registered are limited by the detection scheme. For this reason, we will treat the broadband and narrowband regimes as two separate problems. For both detection regimes we find the rates of $(n-1)$ photon generation via an $n^{\text{th}}$-order process and an $(n-1)$-order process and derive a relationship between these rates.

\subsection{\label{broadbandprocesses} Broadband detection}
Integrating Eq.(\ref{eq:totalcountrateexp}) over all wavevectors captured by the detection bandwidths $\text{D}_i$ gives the $(n-1)$ photon flux into $(n-1)$ detector bandwidths
\begin{multline}
    N_{n-1}^{(n)}=\frac{2\pi}{\hbar^2}\left[\gamma^{(n)}\right]^2  \frac{V_q^n}{(2\pi)^{3(n-1)}} V \times \\
    \int_{\text{D}_1} \cdots \int_{\text{D}_{n-1}}\int_{-\infty}^{\infty} \delta(\Delta \vec{k}^{(n)}) \delta(\Delta \omega^{(n)}) \prod_{i=1}^n \abs{c_i} d\vec{k}_i. \label{eq:bbcountrate}
\end{multline}
The ratio of $(n-1)$ photon generation rates via an $n^{\text{th}}$-order process and an $(n-1)$ order process is given by
\begin{equation}
     \frac{N_{n-1}^{(n)}}{N_{n-1}^{(n-1)}}=\frac{n^2}{32 \pi^3}\frac{\left[\chi^{(n)}\right]^2}{\left[\chi^{(n-1)}\right]^2} \langle E_{bb}^2\rangle,
\end{equation}
where we introduce the squared effective broadband vacuum field \cite{chekhova2005spectral} as
\begin{equation}\label{eq:bbfield}
    \langle E_{bb}^2\rangle= \frac{\hbar}{2 \epsilon_0 c}\int_{-\infty}^{\infty} \frac{ \omega_n v_n}{ n_n} \xi(\vec{k}_n) d\vec{k}_n,
\end{equation}
where the reduced phase matching function, 
\begin{multline}
    \xi(\vec{k}_n)= \\
    \frac{\int_{\text{D}_1} \cdots \int_{\text{D}_{n-1}} \delta(\Delta\vec{k}^{(n)}) \delta(\Delta\omega^{(n)}) \prod_{i=1}^{n-1} \abs{c_i} d\vec{k}_i}{\int_{\text{D}_1} \cdots \int_{\text{D}_{n-1}} \delta(\Delta\vec{k}^{(n-1)}) \delta(\Delta\omega^{(n-1)}) \prod_{i=1}^{n-1} \abs{c_i} d\vec{k}_i},
\end{multline}
accounts for the increased number of ways to fulfill the phase matching condition with $n$ modes.%In the coming sections we will refer to the combination of the frequency and angles that satisfy exact phase matching as the frequency-angular contour. 
The effective broadband vacuum field is the total electric field of all photons in mode $\vec{k}_n$ that satisfy the reduced phase matching condition. 
\subsection{\label{narrowbandprocess} Narrowband detection \label{nbdetec}}
The rate of $(n-1)$ photon generation into $n-1$ narrow detection intervals, $\text{D}_i$, is
\begin{multline}
    \Delta N^{(n)}_{n-1}=\frac{2\pi}{\hbar^2}\left[\gamma^{(n)}\right]^2  \frac{V_q^n}{(2\pi)^{3n}} \times \\
    \int_{-\infty}^{\infty} D(\Delta \vec{k}^{(n)}, \Delta \omega^{(n)}) \abs{c_n} d\vec{k}_n \prod_{i=1}^{n-1} \abs{c_i} \text{D}_i.
\end{multline}
The ratio of the $(n-1)$ photon generation rate for an  $n^{th}$-order process and an $(n-1)$-order process is
\begin{equation}
    \frac{\Delta N^{(n)}_{n-1}}{\Delta N^{(n-1)}_{n-1}}=\frac{n^2}{32 \pi^3}\frac{\left[\chi^{(n)}\right]^2}{\left[\chi^{(n-1)}\right]^2} \langle E_{nb}^2\rangle,
\end{equation}
where we define the squared effective narrowband vacuum field as 
\begin{equation}\label{eq:narrowbandfield}
    \langle E^2_{nb} \rangle= \frac{\hbar}{2 \epsilon_0 c V^2}\int_{-\infty}^{\infty} \frac{\omega_n v_n}{n_n } \abs{f\left(\Delta \vec{k}^{(n)})\right)}^2 d\vec{k}_n,
\end{equation}
assuming that the $(n-1)$-order process is exactly phase matched and the $n^{\text{th}}$-order process satisfies energy conservation.

Approximating the non-linear susceptibility as $\chi^{(n)}\approx E_{a}^{-n}$, where $E_{a}$ is the atomic field strength \cite{boyd2003nonlinear}, one finds that the ratio of the effective vacuum field and atomic field gives the reduction in efficiency from a high order process to the next lower order process:
\begin{equation}
    \frac{N^{(n)}_{n-1}}{N^{(n-1)}_{n-1}}= \frac{n^2}{32 \pi^3}\frac{\langle E_{bb}^2\rangle}{E_a^2},
\end{equation}

\begin{equation}
    \frac{\Delta N^{(n)}_{n-1}}{\Delta N^{(n-1)}_{n-1}}= \frac{n^2}{32 \pi^3}\frac{\langle E_{nb}^2\rangle}{E_a^2}.
\end{equation}

\section{\label{sec:seed}Seeding}

In this section we move from the spontaneous generation of photons via $n^{th}$-order SPDC, to the case where we stimulate the process using a coherent seed beam. We assume that the seed has a wavevector $\vec{k}_s=\vec{k}_n$ and a frequency $\omega_s=\omega_n$. If the seed has a large intensity then a classical field description is adequate. The Hamiltonian of a seeded process is therefore
\begin{equation}
    \hat{H}^{(n)}_s=\gamma_s^{(n)}  f_s(\Delta \vec{k}^{(n)}) (\imath)^{n-1}\prod_i^{n-1}\sqrt{c_i} a_i^{\dagger}+h.c.,
		\label{eq:Hams}
\end{equation}
where \cite{note1},
\begin{equation}\label{eq:gammas}
    \gamma_s^{(n)}=-\frac{n!\chi^{(n)}}{2^n}\sqrt{\frac{4 I_s I_p}{c^2 n_p n_s}},
\end{equation}
$I_s$ is the seed intensity and the phase matching function in the seeded case is 
\begin{equation}
     f_s(\Delta \vec{k}^{(n)})=\int F(\vec{r}) A_p(x,y) A_s(x,y) e^{\imath \Delta \vec{k} \cdot \vec{r}} d^3\vec{r},
		\label{eq:phasematch_s}
\end{equation}
where $A_s(x,y)$ is the transverse field distribution of the seed beam.

The seeded $n^{\text{th}}$-order Hamiltonian (\ref{eq:Hams}) contains $(n-1)$ photon creation operators. This means that the characteristics of $n$ photon emission are lost and the photon statistics become similar to that of $(n-1)$ photon SPDC. 

The rate of transition to an $(n-1)$ photon state is
\begin{equation}
    \Gamma_s^{(n)}=\frac{2\pi}{\hbar^2}\left[\gamma_s^{(n)}\right]^2 D(\Delta \vec{k}^{(n)}, \Delta \omega^{(n)}) \prod_i^{n-1}\abs{c_i},
\end{equation}
which gives the rate of transition into the intervals $d\vec{k}_i$ 
\begin{multline}
    \label{eq:seed}
    dN_{s,n-1}^{(n)}= \\ \frac{2\pi}{\hbar^2}\left[\gamma_s^{(n)}\right]^2 D(\Delta \vec{k}^{(n)}, \Delta \omega^{(n)}) \frac{V_q^{n-1}}{(2\pi)^{3(n-1)}} \prod_{i=1}^{n-1} \abs{c_i} d\vec{k}_i.
\end{multline}

\subsection{Broadband detection}
The seeded $(n-1)$ photon emission rate in the broadband case is  
\begin{multline}
    N_{s,n-1}^{(n)}=\frac{2\pi}{\hbar^2}\left[\gamma_s^{(n)}\right]^2 \frac{V_q^{n-1}}{(2\pi)^{3(n-1)}} \times \\ 
    \int_{\text{D}_1} \cdots \int_{\text{D}_{n-1}} D(\Delta \vec{k}^{(n)}, \Delta \omega^{(n)}) \prod_{i=1}^{n-1} \abs{c_i} d\vec{k}_i.
	\label{eq:bbseed}
\end{multline}
By taking the ratio of Eq.(\ref{eq:bbcountrate}) and (\ref{eq:bbseed}) we obtain
\begin{equation}\label{eq:bbseedratio}
    \frac{ N_{s,n-1}^{(n)}}{N_{n-1}^{(n)}}= \frac{|E_s|^2}{\langle E_{bb}^2\rangle},
\end{equation}
where the squared seed field amplitude is
\begin{equation}
     |E_s|^2=\frac{2 I_s}{\epsilon_0 n_s c}.
\end{equation}

\subsection{Narrowband detection}
The seeded $(n-1)$ photon emission rate into $(n-1)$ narrowband detectors with bandwidths $\text{D}_i$ is
\begin{multline}
     \Delta Ns_{n-1}^{(n)}= \frac{2\pi}{\hbar^2}\left[\gamma_s^{(n)}\right]^2 \frac{V_q^{n-1}}{(2\pi)^{3(n-1)}} \times\\ 
     D(\Delta \vec{k}^{(n)}, \Delta \omega^{(n)}) \prod_{i=1}^{n-1} \abs{c_i} \text{D}_{i}.
	\label{eq:nbseed}
\end{multline}
    
The ratio of the $(n-1)$ photon emission rate into $(n-1)$ detectors for a seeded and unseeded $n^{th}$-order process is
\begin{equation}\label{eq:nbseedratio}
    \frac{\Delta Ns_{n-1}^{(n)}}{\Delta N_{n-1}^{(n)}}= \frac{|E_s|^2}{\langle E_{nb}^2\rangle}.
\end{equation}

In both broadband and narrowband cases the use of a seed has an advantage only if the seed field is larger than the corresponding effective vacuum field.

In the broadband case if the reduced phase matching function, $\abs{\xi(\vec{k}_n)}$, is broad then the effective field is comparatively large. Using a continuous wave seed would require too high intensities to overcome the effective broadband vacuum field strength. In this situation it is advantageous to work with pulsed seed and pump. The product of the seed and pump peak power averaged over time yields a factor of inverse duty cycle, significantly enhancing the efficiency of the $n^{th}$-order process. For the seed to be used optimally one must overlap the pump and seed waves in space and time. This follows from the phase matching function Eq.~(\ref{eq:phasematch_s}), which is given by the convolution of the seed and pump. 

\section{Efficiency of unseeded TOPDC}\label{sec:unseededcalc}
Here we consider the case of $n=3$, which corresponds to TOPDC. We estimate the photon emission rates for both seeded and unseeded TOPDC in rutile (\ce{TiO_2}), which has been suggested previously as potentially efficient for TOPDC \cite{borshchevskaya2015three}.

From Eq.(\ref{eq:totalcountrateexp}), the differential generation rate of a three-photon state via TOPDC into modes $k_1$, $k_2$ and $k_3$ is
\begin{multline}\label{eq:TOPDCEFF}
    \mathscr{N}(\vec{k}_{1}, \vec{k}_{2}, \vec{k}_{3})=\frac{\,dN^{(3)}}{\,d\vec{k}_{1}\,d\vec{k}_{2}\,d\vec{k}_3}=\\ 
    R^{(3)} \frac{\omega_1 \omega_2 \omega_3 v_1 v_2 v_3}{n_1 n_2 n_3} \abs{\tilde {f}(\Delta \vec{k}^{(3)})}^2\delta(\Delta \omega^{(3)}),
\end{multline}
where
\begin{equation}\label{eq:r}
    R^{(3)}=\frac{\hbar V}{8  (2 \pi)^8 \epsilon_0^3 c^3}[\gamma^{(3)}]^2.
\end{equation}

Again we consider the cases of 1) broadband detection and 2) narrowband detection. In both situations we can chose to detect the rate of triple photons in modes $k_1$, $k_2$ and $k_3$, the rate of double photons in modes $k_1$, $k_2$ or the rate of single photons in mode $k_1$. For double and single photon count rates Eq.~(\ref{eq:TOPDCEFF}) is integrated over all wavevectors of the unregistered mode (mode 3).

\subsection{Broadband detection of TOPDC}
Following from section~\ref{sec:phasematching}, the phase matching function in Eq.~(\ref{eq:TOPDCEFF}) can be replaced by a delta function. Calculating the integrals
\begin{subequations} \label{eq:bbrate}
    \begin{align}
        N_{3} &= \int_{\text{D}_1}\int_{\text{D}_2}\int_{\text{D}_3}\mathscr{N}(\vec{k}_{1}, \vec{k}_{2}, \vec{k}_{3})\,d\vec{k}_{1}\,d\vec{k}_{2}\,d\vec{k}_3, \label{eq:bbratesub1} \\
        N_{2} &= \int_{-\infty}^{+\infty}\int_{\text{D}_2}\int_{\text{D}_3}\mathscr{N}(\vec{k}_{1}, \vec{k}_{2}, \vec{k}_{3})\,d\vec{k}_{1}\,d\vec{k}_{2}\,d\vec{k}_3, \label{eq:bbratesub2} \\
        N_{1} &= \int_{-\infty}^{+\infty}\int_{-\infty}^{+\infty}\int_{\text{D}_3}\mathscr{N}(\vec{k}_{1}, \vec{k}_{2}, \vec{k}_{3})\,d\vec{k}_{1}\,d\vec{k}_{2}\,d\vec{k}_3 \label{eq:bbratesub3}
    \end{align}
\end{subequations}
gives the triple photon, double photon, and single photon fluxes into three, two and one broadband detectors whose bandwidths are given by $\text{D}_1$, $\text{D}_2$ and $\text{D}_3$.

\subsection{Narrowband detection of TOPDC}

The phase matching function in Eq.~(\ref{eq:TOPDCEFF}) is replaced by  Eq.~(\ref{eq:phasematchingGaussianbeam}) in the narrowband regime. The triple, double and single photons fluxes into three, two and one narrowband detectors are given by 
\begin{subequations}\label{eq:nbrate}
\begin{align}
    \Delta N_3 &= \mathscr{N}(\vec{k}_{01}, \vec{k}_{02}, \vec{k}_{03}) \text{D}_1 \text{D}_2 \text{D}_3, \label{eq:nbratesub1}\\
    \Delta N_2 &=\int_{-\infty}^{+\infty} \mathscr{N}(\vec{k}_{1}, \vec{k}_{02}, \vec{k}_{03})\,d\vec{k}_1 \text{D}_2 \text{D}_3,\label{eq:nbratesub2} \\
    \Delta N_1 &= \int_{-\infty}^{+\infty} \int_{-\infty}^{+\infty} \mathscr{N}(\vec{k}_{1}, \vec{k}_{2}, \vec{k}_{03})\,d\vec{k}_{1}\,d\vec{k}_2\text{D}_3, \label{eq:nbratesub3}
\end{align}
\end{subequations}
where $\text{D}_1, \text{D}_2, \text{D}_3$ are the detector bandwidths centered around modes $\vec{k}_{01}, \vec{k}_{02}, \vec{k}_{03}$.

In practice integration over all unregistered wavevectors is impossible as dispersion information only exists for a limited bandwidth. In the calculations below we integrate over a range of wavevectors where dispersion relation for \ce{TiO_2} still holds true, which results in an \textit{underestimation} of the total photon flux. The deviation from the true emission rate is comparatively small, due to the factor $\omega_1\omega_1\omega_3$ in Eq.~(\ref{eq:TOPDCEFF}), which implies that near-degenerate frequencies contribute more than non-degenerate frequencies. An additional source of error is introduced by assuming the cubic susceptibility to be independent of frequency. This in general is not true, however \ce{TiO_2} is optically transparent over the frequencies we consider, implying a fairly constant cubic susceptibility. 

From here on in we choose to work in frequency-angle variables ($\omega,\theta,\phi$) as opposed to wavevector space ($\vec{k}$), the transformation is given in Appendix \ref{sec:appendix1}. The parameter (see Eq.~(\ref{eq:TOPDCEFF}))
\begin{eqnarray}\label{eq:intparam}
    \int_{-\infty}^{+\infty} \int_{-\infty}^{+\infty}   \mathscr{N} (\vec{k}_{1}, \vec{k}_{2}, \vec{k}_{3})\,d\vec{k}_{1}\,d\vec{k}_2= (2\pi)^4 R^{(3)} \times\nonumber\\
		\frac{\omega_3  v_{3} }{c^2 n_3}
		\int \frac{\tilde{\omega}_1(\theta_2,\vec{k}_3) \tilde{\omega}_2^3(\theta_2,\vec{k}_3) \tilde{v}_1\tilde{n}_2}{\tilde{n}_1^2 } \sin(\theta_2) d\theta_2
\end{eqnarray}
is common to both the narrowband and broadband single photon emission rates and gives information on the spread and spectral content of the photons emitted. The frequency $\tilde{\omega}_1(\theta_2,\vec{k}_3)$ is given by Eq.~(\ref{eq:omega1}) and $\tilde{v}_1$, $\tilde{n}_1$ are the group and refractive index evaluated at $\tilde{\omega}_1(\theta_2,\vec{k}_3)$. The frequency $\tilde{\omega}_2(\theta_2,\vec{k}_3)$ is given in Eq.~(\ref{eq:pmcondition}), $\tilde{n}_2$ is the refractive index evaluated at this frequency, and $\theta_2$ is the polar angle of mode 2.

\begin{figure}[htbp]
    \includegraphics[width=8cm]{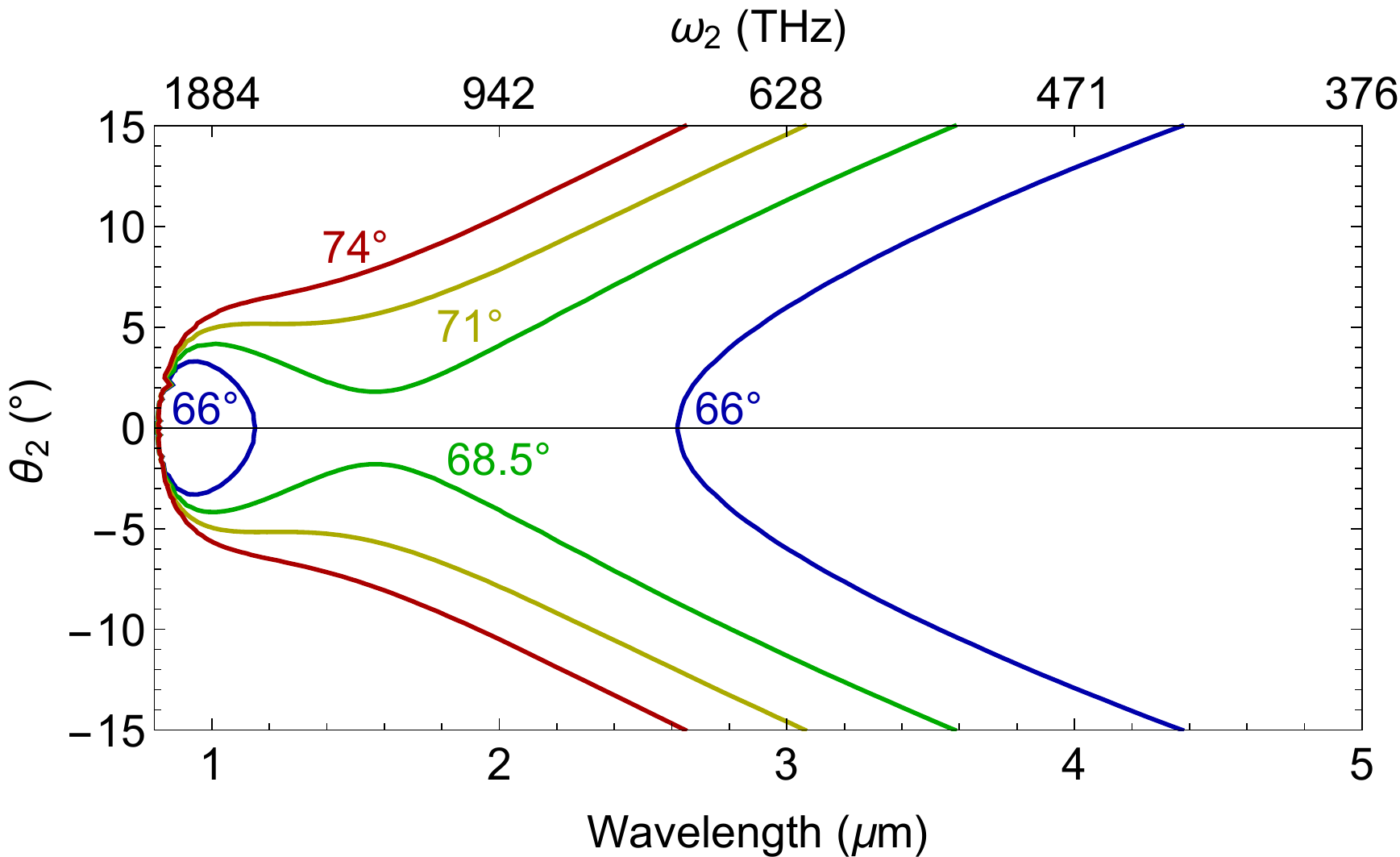}
\caption{ Frequency-angle contour for a \ce{TiO_2} crystal with type-I (o $\rightarrow$ eee) phase matching at different orientation angles. The pump wavelength is $\lambda_p=\SI{532}{\nano\meter}$ and the parameters for mode 3 are fixed to $\lambda_3=\SI{1596}{\nano\meter}$ and $\theta_3=0^{\circ}$.}
\label{fig:pmcurves}
\end{figure}

A key point in all further calculations is finding the dependence $\tilde{\omega}_2(\theta_2,\vec{k}_3)$ at fixed $\vec{k}_3$ (assuming azimuthal symmetry). This dependence, which we will refer to as the frequency-angle contour of mode 2, gives all the points $\omega_2,\theta_2$ that satisfy conditions $\Delta \vec{k}^{(3)}=0, \Delta \omega^{(3)}=0$ at fixed $\vec{k}_3$. An example is plotted in Fig.~\ref{fig:pmcurves} for TOPDC in \ce{TiO_2} assuming the pump wavelength to be $\lambda_p=\SI{532}{\nano\meter}$ and mode 3 parameters fixed to collinear degenerate case: $\lambda_3=\SI{1596}{\nano\meter}$, $\theta_3=0^{\circ}$.

Phase matching is satisfied using the birefringence present in \ce{TiO_2} crystals. Type-I phase matching (o$\rightarrow$eee) is assumed. Varying the angle subtended by the optic axis of the crystal and the pump propagation direction, further called the crystal orientation, it is possible to change the shape of the frequency-angle contour as shown in Fig.~\ref{fig:pmcurves}. 

\begin{figure}[htbp]
  \includegraphics[width=8cm]{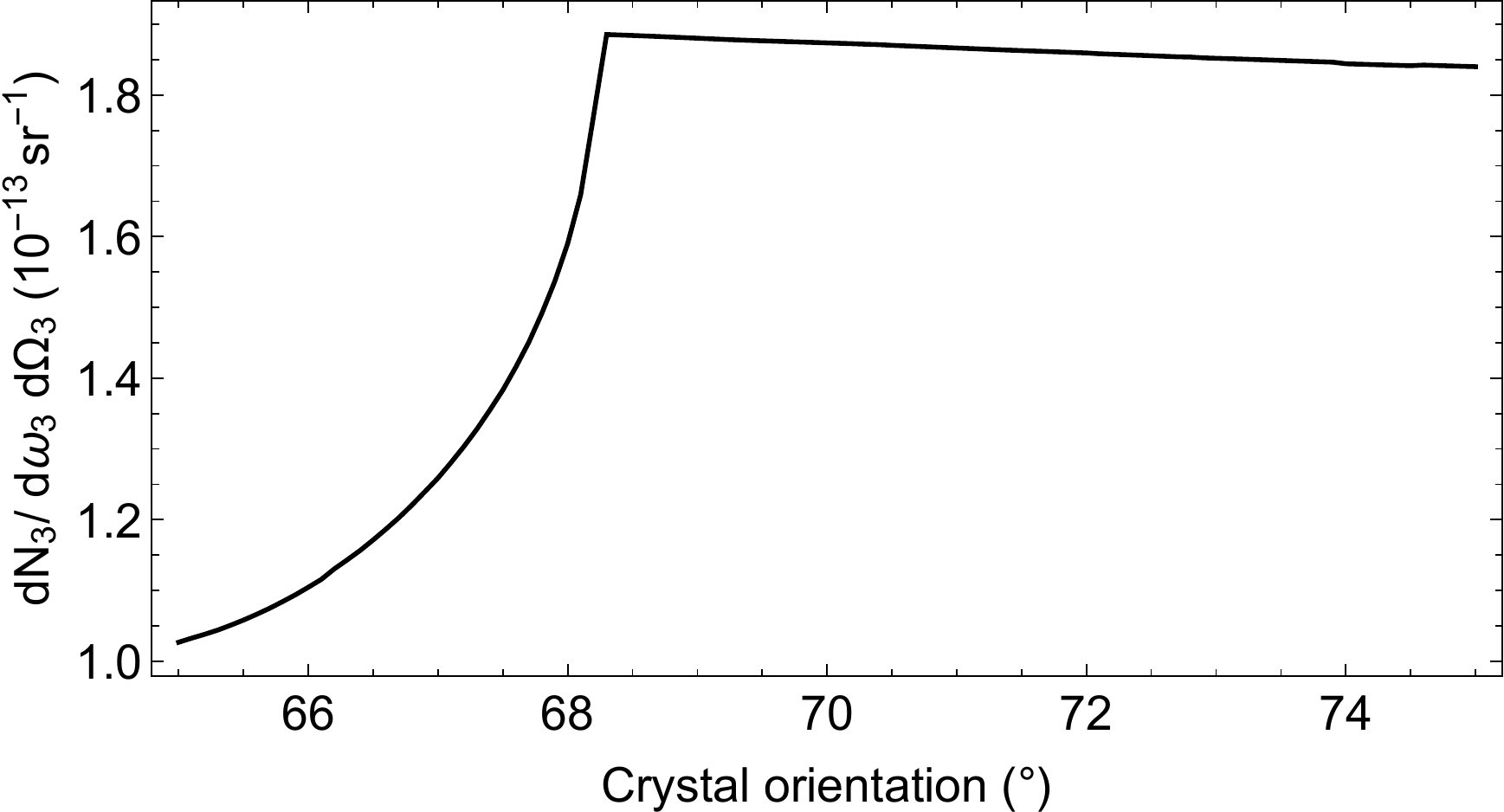}
\caption{The differential single photon count rate for TOPDC in a $5$ mm \ce{TiO_2} crystal pumped by $100$ mW as a function of the crystal orientation angle. The pump wavelength is $\lambda_p=\SI{532}{\nano\meter}$ and the parameters for mode 3 are $\lambda_3=\SI{1596}{\nano\meter}$ and $\theta_3=0^{\circ}$.}
\label{fig:pmefflength}
\end{figure}

The value of Eq.(\ref{eq:intparam}) and by extension the single photon flux scales as the {\it length} of the frequency-angle contour. How the single photon differential rate (Eq.(\ref{eq:intparam})) changes with the crystal orientation is plotted in Fig.~\ref{fig:pmefflength}. The highest rate occurs at the crystal orientation $68.24^{\circ}$, corresponding to a contour that crosses the collinear degenerate point; hence for all subsequent calculations this orientation was considered. The discontinuity in Fig.~\ref{fig:pmefflength} is due to the fact that the phase matching width scales as $L^{-1/2}$ at the degeneracy point and not as $L^{-1}$, which is assumed when we replace the phase matching function with a delta function.  

\begin{figure}
       \includegraphics[width=9cm]{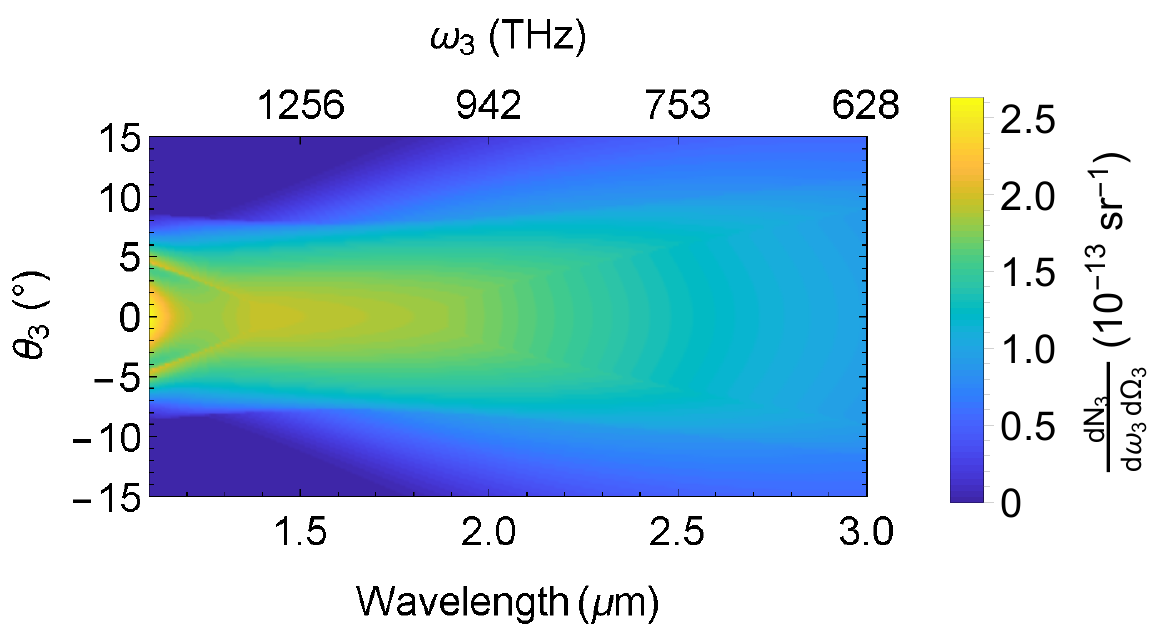}
    \caption{The frequency-angle spectrum of single-photon differential generation rate in mode 3 for unseeded TOPDC in \ce{TiO_2} crystal oriented at $68.24^{\circ}$. The crystal length $L=5$ mm and the pump power $P_p=\SI{100}{\milli\watt}$.}\label{fig:angularfeqspec} 
\end{figure}

Plotting the integrated value of the frequency-angular contour as a function of the mode 3 parameters gives the frequency-angle spectrum of the single photon emission from unseeded TOPDC, shown in Fig.~\ref{fig:angularfeqspec}.
The resulting spectrum is very broad compared to the equivalent spectrum for SPDC. This is expected in TOPDC, due to the increased number of degrees of freedom in which to satisfy phase matching. In fact we expect this trend to continue when looking at higher orders of SPDC. Nevertheless, the TOPDC spectrum is not uniform and, for a fixed frequency, it does not span all angles. It follows from Eq.(\ref{eq:bbratesub3}) that integration over a region of the spectrum shown in Fig.~\ref{fig:angularfeqspec}, with the limits set by $\text{D}_3$, gives the total singles emission rate $N_1$ for the broadband case.

\section{Efficiency of singly seeded TOPDC}\label{sec:seededcalc}
From the general consideration of Sec.~\ref{sec:seed}, singly seeded TOPDC results in the emission of photon pairs, similar to two-photon SPDC. This is a consequence of fixing one of the final states in TOPDC. In this case we fix mode 3 such that $\vec{k}_3=\vec{k}_s$. From Eq.~(\ref{eq:seed}) the differential rate of two photon transitions via seeded TOPDC is 
\begin{equation}\label{eq:seedTOPDC}
    \mathscr{N}_s=\frac{\,dN_s^{(3)}}{\,d\vec{k}_1\,d\vec{k}_2}= R_s^{(3)} \frac{\omega_1 \omega_2 v_1 v_2}{n_1 n_2} \abs{\tilde{f}(\Delta \vec{k}_s^{(3)})}^2\delta(\Delta \omega_s^{(3)}),
\end{equation}
where 
\begin{equation}
    R_s^{(3)}=\frac{V}{4(2\pi)^5 \epsilon_0^2 c^2}[\gamma_s^{(3)}]^2.
\end{equation}
 
\subsection{Broadband detection of seeded TOPDC}
Eqs.~(\ref{eq:bbrate}) and (\ref{eq:nbrate}) are simplified as seeded TOPDC only generates pair states; hence we are limited to detecting just photon pairs and single photons. The rate of two-photon and single photon states collected by two and one detectors, respectively, is
\begin{subequations}
\begin{align}\label{eq:seededbbrate}
    N_{s,2} &= \int_{\text{D}_1}\int_{\text{D}_2} \mathscr{N}_s(\vec{k}_1,\vec{k}_2)\,d\vec{k}_{1}\,d\vec{k}_{2},\\
    N_{s,1} &= \int_{\text{D}_1}\int_{-\infty}^{+\infty}\mathscr{N}_s(\vec{k}_1,\vec{k}_2)\,d\vec{k}_{1}\,d\vec{k}_{2}.
\end{align}
\end{subequations}

\subsection{Narrowband detection of seeded TOPDC}

The rate of double and single photon emission into narrow bands $\text{D}_{2}$ and $\text{D}_{1}$ is
\begin{subequations}
\begin{eqnarray}\label{eq:seededbbrate}
     \Delta N_{s,2}= \mathscr{N}_s(\vec{k}_{01},\vec{k}_{02}) \text{D}_1 \text{D}_2, \\
    \Delta N_{s,1}= \int_{-\infty}^{+\infty} \mathscr{N}_s(\vec{k}_1,\vec{k}_{02})\,d\vec{k}_{1}\text{D}_2.
\end{eqnarray}
\end{subequations}
where $\vec{k}_{01}$ and $\vec{k}_{02}$ represent the central wavevectors of the detection bands $\text{D}_2$ and $\text{D}_1$. 

The seeded analogue of  Eq.~(\ref{eq:intparam}) is
\begin{multline}\label{eq:diffseedmain}
      \int_{-\infty}^{+\infty}\mathscr{N}_s (\vec{k}_1,\vec{k}_2)\,d\vec{k}_{1}= 2\pi R_s^{(3)} V \frac{\tilde{\omega}_1^3\omega_2 n_1 v_2}{n_2} \times\\ \exp \left(-\frac{\Delta k_x^2+\Delta k_y^2}{4}w_0^2\right)
    \sinc^2\left(\frac{\Delta k_z L}{2}\right),
\end{multline}
which, again, is common to both the single broadband and single narrowband seeded transition rates. For the sake of completeness we assume that the phase matching function can be broader than $\text{D}_1$, hence in Eq.~(\ref{eq:diffseedmain}) the phase matching function is given by Eq.(\ref{eq:phasematchingGaussianbeam}).

\begin{figure}
   \includegraphics[width=9cm]{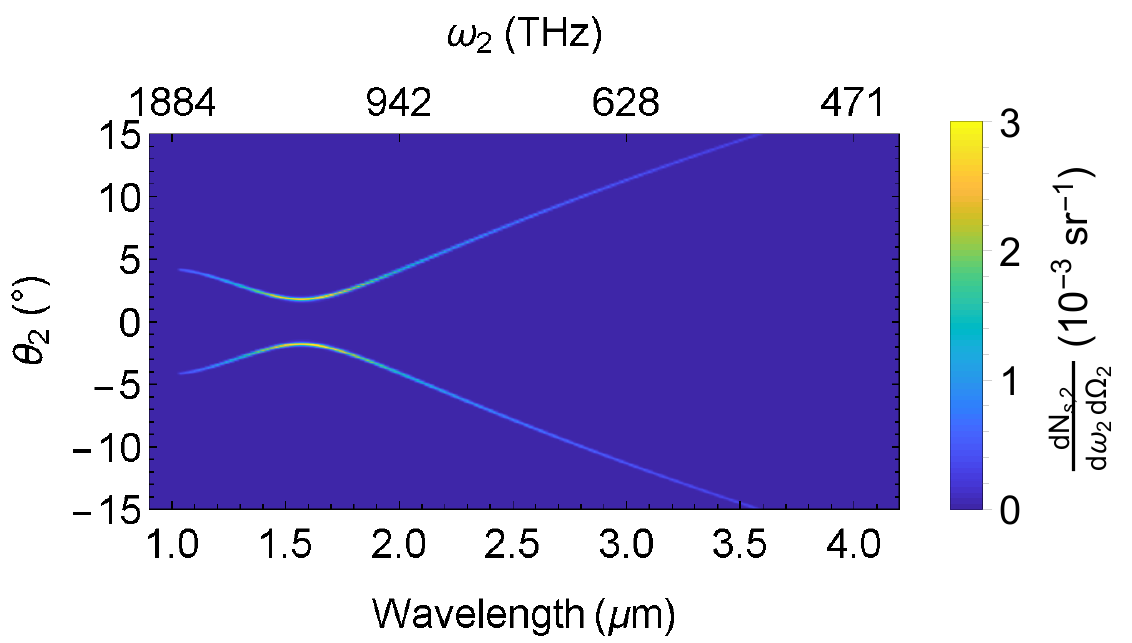}
    \caption{The frequency-angle  spectrum of mode 2 for seeded TOPDC in \ce{TiO_2} of length $L$= $\SI{5}{\milli\meter}$, for the pump power $P_p=\SI{100}{\milli\watt}$ and the crystal orientation $68.5^{\circ}$.}
		\label{fig:seedangularfeqspec} 
\end{figure}

The differential single photon emission rate is plotted in Fig.~\ref{fig:seedangularfeqspec} as a function of the frequency/wavelength and angle of emission in mode 2. The first noticeable feature is that the  frequency-angle spectrum resembles the one typically observed in SPDC. This follows from our earlier statement that seeding an $n^{\text{th}}$-order process reduces the $n^{\text{th}}$-order Hamiltonian to a $(n-1)$-order Hamiltonian. The second is that when $\vec{k}_3=\vec{k}_s$ the frequency-angle contour given by $\tilde{\omega}_2(\Omega_2,\vec{k}_s)$  is similar to the spectrum in Fig.~\ref{fig:seedangularfeqspec}. The difference between the two cases is that the  frequency-angle spectrum has a non-zero width due to the non-zero width of the phase matching function. A significant point is that the integrated values of both the frequency-angular contour in Fig.~\ref{fig:angularfeqspec} and the frequency-angular spectrum in Fig.~\ref{fig:seedangularfeqspec} over all frequency and angle space are equivalent.

Comparing the frequency-angle spectrum of seeded TOPDC in Fig.~\ref{fig:seedangularfeqspec} and the equivalent spectrum for spontaneous TOPDC shows that the seeded spectrum is far more concentrated to particular regions. This is advantageous, especially in the narrowband case, as it is clear where to place detectors to collect the largest number of single photons. Conversely the TOPDC spectrum is broad and uniform over a large region of the angular frequency spectrum, making it harder to distinguish from background sources of light.  

Comparing Eq.~(\ref{eq:intparam}) for spontaneous TOPDC and Eq.~(\ref{eq:diffseedmain}) for seeded TOPDC, reveals that the reduced Planck constant enters only Eq.~(\ref{eq:intparam}), through $R^{(3)}$, and is absent from Eq.~(\ref{eq:diffseedmain}).
It is also absent for the equivalent relations for SPDC. This, along with other effects such as a non-Gaussian Wigner function \cite{banaszek1997quantum,elyutin1990three}, implies that spontaneous TOPDC displays quantum features that are not observed in either seeded TOPDC or SPDC. More generally, it follows from Eq.~(\ref{eq:totalcountrateexp}) that SPDC of order $n$ contains the Planck constant to the power $n-2$.

\section{Numerical estimations of seeded and unseeded photon count rates}\label{sec:estimates}

In this section we take into account the available experimental parameters, including the quantum efficiencies $\eta_{1,2,3}$ for the three detectors, and estimate the count rates for triple coincidences $\eta_1 \eta_2 \eta_3 N_3$, double coincidences $\eta_1 \eta_2 N_2$ and single photon counts $\eta_1 N_1$ from unseeded TOPDC in both the broadband regime and the narrowband regime. The quantum efficiencies are assumed to be independent of frequency or angle. We take the effective cubic susceptibility for the type-I  process in \ce{TiO_2} to be $\chi^{(3)}=\SI{2.1E-20}{\meter\squared\per\volt\squared}$ \cite{boyd2003nonlinear}, the length $L=\SI{5}{\milli\meter}$ and the orientation angle $68.24^{\circ}$ (see Fig.\ref{fig:pmefflength}).
For spontaneous TOPDC we assume a CW pump with power $P_p=\SI{100}{\milli\watt}$. For seeded TOPDC we assume a pulsed pump and seed (see Section.~\ref{sec:seed}) overlapped with a beam waist $w_0=\SI{100}{\micro\meter}$ and duty cycle $\mathscr{D}=\SI{2E-8}{}$. The seed has mean power $P_s=\SI{10}{\milli\watt}$ and the pump, $P_p=\SI{100}{\milli\watt}$ which corresponds to peak intensities of $I_s=\SI{1.6E13}{\watt\per\meter\squared}$ and $I_p=\SI{1.6E14}{\watt\per\meter\squared}$, respectively. The seed wavelength is assumed to be $1620\ $nm, slightly red-shifted from the detection bandwidth.

\subsection{Broadband detection}
Broadband detection refers to both frequency and angle. In this situation the best strategy is using multimode APDs \cite{ID220}. Despite their lower quantum efficency compared to superconducting nanowires, they are more efficient to use due to the larger collection angle. For such a situation we assume a quantum efficiency of ~15\% over a frequency range of $\SI{1200}{\nano\meter}$ to $\SI{1600}{\nano\meter}$. The capture angle is limited only by the aperture size of the emission collected which, for arguments sake, we limit to $-10^{\circ}$ to $10^{\circ}$. 

\begin{figure}[htbp]
   \includegraphics[width=8.5cm]{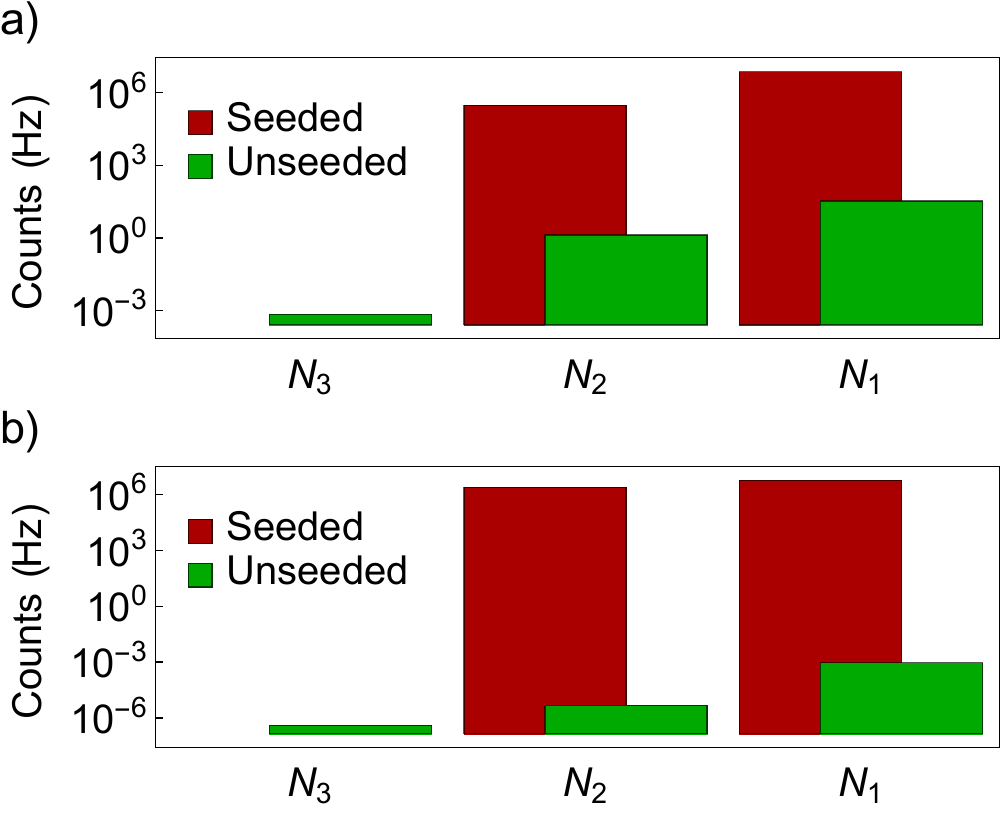}
\caption{The expected triple, double and single photon count rates of seeded and unseeded TOPDC in $\SI{5}{\milli\meter}$ \ce{TiO_2}   for a) the broadband regime and b) the narrowband regime. The seed beam is centered at 1620 nm, the pump power is $P_p=\SI{100}{\milli\watt}$ and the overlapped seed peak intensity $I_s=\SI{1.6E13}{\watt\per\meter\squared}$. The count rates calculated for the broadband regime assume  collection of wavelengths from $\SI{1200}{\nano\meter}$ to $\SI{1600}{\nano\meter}$ and angles from $-10^{\circ}$ to $10^{\circ}$ with a detection efficiency of 15\%. The count rates for the narrowband regime assume  collection of wavelengths from $1579$ nm to $1589$ nm and angles from $-0.25^{\circ}$ to $0.25^{\circ}$ with a detection efficiency of 80\%.}
\label{fig:totcounts}
\end{figure}

Figure \ref{fig:totcounts}.~a) shows the expected triple coincidence, double coincidence, and single count rates via seeded and unseeded TOPDC in \ce{TiO_2}. Without seeding, it is feasible to observe single counts and double coincidences whose rates are 33 Hz and 1 Hz, respectively. The estimated triple coincidence rate is on the order of a few per hour and would be more difficult to measure experimentally. 

The seed enhances the detection rate of single and double photons by roughly $10^{7}$ times compared to the case of unseeded TOPDC, although this comes at the cost of losing the three-photon state. The seeded two photon emission rate is $\SI{300}{\kilo\hertz}$. With such rates one can readily study the properties of the three-photon state generated via spontaneous TOPDC by using methods such as stimulated emission tomography.

\subsection{Narrowband detection} 
Working in the narrowband detection regime is advantageous if fluorescence or other sources of noise are competing processes. By choosing a bandwidth where the three photon emission is particularly strong one can maximize the signal to noise ratio. Emission rates around the degeneracy frequency are relatively high due to the factor  $\omega_1 \omega_2 \omega_3$ that appears in Eq.~(\ref{eq:transpermode}). For a $\SI{100}{\micro\meter}$ beam waist, single-mode superconducting nanowire detectors capture angles between $-0.25^{\circ}$ and $0.25^{\circ}$ with a quantum efficiency of 80\%. We assume that the frequency bandwidth is restricted to $\SI{1584}{\nano\meter}\pm\SI{5}{\nano\meter}$, which is close to degeneracy but satisfies the frequency matching in the seeded case.

The calculated narrowband triple, double and single photon count rates are compared in Fig.~\ref{fig:totcounts}~b). Here the rates are lower than in the broadband case, but the seed provides a stronger enhancement.  This is expected as seeding fixes one of the triplet modes thereby reducing the number of degrees of freedom; it follows that photons are emitted into a fewer number of modes but the count rate per mode increases. %The increase in seeded double and single photon count rates in the narrowband regime corresponds to the ratio given in Eq.~(\ref{eq:nbseedratio}) which is around $\SI{1.4E14}{}$. 

\section{Conclusion}\label{sec:conc}

In conclusion, we have derived a general expression for the emission rate of $n^{\text{th}}$-order SPDC. The phase matching and energy conservation conditions can be written as a single function $D(\Delta \vec{k}^{(n)}, \Delta \omega^{(n)})$, which restricts the  frequency-angle distribution. As the function $D(\Delta \vec{k}^{(n)}, \Delta \omega^{(n)})$ restricts only two degrees of freedom, the angular spectrum of two-photon SPDC is fully defined. However moving to higher-order SPDC the number of degrees of freedom exceeds the restrictions set by $D(\Delta \vec{k}^{(n)}, \Delta \omega^{(n)})$ leading to broader angular spectra with increasing $n$.

Calculations are simplified by distinguishing between two regimes, with broadband and narrowband collection of $n$ photon radiation. In the broadband regime the width of the distribution $D(\Delta \vec{k}^{(n)}, \Delta \omega^{(n)})$ is irrelevant as only the integrated distribution matters. As such, the TOPDC efficiency does not depend on the width of the frequency angular spectrum (see Fig.~\ref{fig:seedangularfeqspec} and Fig.~\ref{fig:pmcurves}) for modes 1 and 2, only the length of the curve. In the narrowband case $D(\Delta \vec{k}^{(n)}, \Delta \omega^{(n)})$ is evaluated at the central position of the detector bandwidths. 

By comparing the rate of $(n-1)$-photon state generation for an $n^{\text{th}}$-order process and an $(n-1)$-order process, one can conveniently define an effective field that describes the total electric field of all phase matched states. We show that the ratio of this effective field squared and the atomic field squared roughly gives the reduction in efficiency from one process to the next.  

The rate of emission for $n^{\text{th}}$-order SPDC scales as $\hbar^{n-2}$. Scaling with the Planck constant is a feature commonly attributed to quantum characteristics. This suggests that SPDC of higher orders can be considered less classical than two-photon SPDC. This is in agreement with the fact that these processes produce non-Gaussian states while the output state of second-order SPDC, without post-selection, is Gaussian.

Coherently seeding one of the modes that satisfies phase matching, maps the $n^{th}$-order non-linear Hamiltonian to a  $(n-1)$-order Hamiltonian. As a result, the photon statistics of an $n^{\text{th}}$-order seeded process mimic those of an $(n-1)$-order process. Moreover, using a seed gives an enhancement to the rate of photon emission, equal to the ratio of the seed field amplitude squared and the effective squared vacuum field. When working in the pulsed regime one uses the peak seed field amplitude, hence there can be a large enhancement using a seed.  

Finally, for the particular case of TOPDC ($n=3$) we have estimated the photon emission rates for rutile. We show that despite being far broader than the typical second-order SPDC spectrum, the distribution $D(\Delta \vec{k}^{(n)}, \Delta \omega^{(n)})$ still limits the angular spectrum of TOPDC. Plotting the distribution over frequency and angle reveals optimum regions in which to collect single, double and triplet photons.

The estimated triplet rates are too low to readily observe, however we show that by utilising a pulsed seed beam one can greatly improve the rate of emission. From the estimates the seed gives an enhancement of roughly $10^9$ in the broadband case and $10^{14}$ in the narrowband case. The drawback is that one can only observe double photon counts in the seeded regime and statistics predicted for the three photon state  cannot be acquired. Despite this, using stimulated tomography one can reconstruct the statistics of the three photon state \cite{liscidini2013stimulated}.

\appendix
\section{Derivation of emission rates for unseeded TOPDC \label{sec:appendix1}}

The integrals
\begin{subequations}
\begin{align}
    & \int_{-\infty}^{+\infty}\mathscr{N}(\vec{k}_1,\vec{k}_2)\,d\vec{k}_1,\\
    & \int_{-\infty}^{+\infty}\int_{-\infty}^{+\infty}\mathscr{N}(\vec{k}_1,\vec{k}_2) \,d\vec{k}_{1}\,d\vec{k}_2
\end{align}
\end{subequations}
can be solved analytically. As the integration domain is broader than the phase matching function we can replace $\abs{\tilde{f}\left(\Delta k^{(n)}\right)}^2$ with (\ref{eq:pmdelta}), which gives
\begin{eqnarray}\label{eq:doublediff}
    \int_{-\infty}^{+\infty}\mathscr{N}(\vec{k}_1,\vec{k}_2)\,d\vec{k}_1= (2\pi)^3 R^{(3)} \times\nonumber\\
		\int \frac{\omega_1 \omega_2 \omega_3 v_1 v_2 v_3}{n_1 n_2 n_3}
		\delta(\Delta \vec{k}^{(3)})\delta(\Delta \omega^{(3)}) d\vec{k}_{1}.
\end{eqnarray}
The frequency is related to the wavevector as
\begin{equation}
    \frac{n(\omega_i)\omega_i}{c}=\sqrt{k_{ix}^2+k_{iy}^2+k_{iz}^2},
\end{equation}
where $n(\omega_i)$ is the dispersion dependence.
Then, integration in Eq.(\ref{eq:doublediff}) gives
\begin{multline}\label{eq:exdoublediff}
      \int_{-\infty}^{+\infty} \mathscr{N}(\vec{k}_1,\vec{k}_2)\,d\vec{k}_1= (2\pi)^3 R^{(3)} \frac{\tilde{\omega}_1(\vec{k}_2,\vec{k}_3) \omega_2 \omega_3 \tilde{v}_1 v_2 v_3}{\tilde{n}_1 n_2 n_3} \times \\ \delta(\omega_p-\tilde{\omega}_1(\vec{k}_2,\vec{k}_3)-\omega_2-\omega_3),
\end{multline}
where $\tilde{\omega}_1(\vec{k}_2,\vec{k}_3)$ is found from the equation
\begin{eqnarray}
    \left[\frac{n(\tilde{\omega}_1(\vec{k}_2,\vec{k}_3))\tilde{\omega}_1(\vec{k}_2,\vec{k}_3)}{c}\right]^2=(k_{2x}+k_{3x})^2+\nonumber\\
		+(k_{2y}+k_{3y})^2+(k_{2z}+k_{3z}-k_p)^2,
		\label{eq:omega1}
\end{eqnarray}
and $\tilde{n}_1$, $\tilde{v}_1$ are the refractive index and group velocity at this frequency (both functions of $\vec{k}_2,\vec{k}_3$).

Finding the expression for the single photon generation rate requires integrating Eq.(\ref{eq:exdoublediff}) in $\vec{k}_2$.
It is now convenient to pass from the wavevector space to the frequency-angle space, hence we use the following transformations to substitute into Eq.(\ref{eq:omega1}): 
\begin{subequations}
\begin{align}
    k_{ix}&=\frac{n_i \omega_i}{c} \sin{\theta_i}\cos{\phi_i}, \\
    k_{iy}&=\frac{n_i \omega_i}{c} \sin{\theta_i}\sin{\phi_i}, \\
    k_{iz}&=\frac{n_i \omega_i}{c} \cos{\theta_i}.
\end{align}
\end{subequations}
The differential $d\vec{k}_i$ is then 
\begin{equation}\label{eq:diffint}
      d\vec{k}_{i}=\frac{n_i^2 \omega_i^2}{c^2 v_i} d\omega_i d\Omega_i, 
\end{equation}
where the solid angle interval is
\begin{equation}\label{eq:solidangle}
     d\Omega_i=\sin\left(\theta_i\right)d\theta_i d\phi_i.
\end{equation}
By assuming azimuthal symmetry we can rewrite Eq.~(\ref{eq:omega1}) as
\begin{multline}
    n^2(\tilde{\omega}_1(\vec{k}_2,\vec{k}_3))\tilde{\omega}_1^2(\vec{k}_2,\vec{k}_3)=\\
		\left(n_2 \omega_2\sin{\theta_2}+n_3 \omega_3\sin{\theta_3}\right)^2+\\
		 +\left(n_2 \omega_2\cos{\theta_2}+n_3 \omega_3\cos{\theta_3}-n_p \omega_p\right)^2.
		\label{eq:omega11}
\end{multline}

Due to the azimuthal symmetry, integration of Eq.~\ref{eq:exdoublediff} in $d\phi_2$ results in a factor $2\pi$. Integration in $\omega_2$ fixes it to be
\begin{equation}\label{eq:pmcondition}
    \tilde{\omega}_2(\omega_3,\theta_2,\theta_3)=\omega_p-\tilde{\omega}_1(\tilde{\omega}_2,\omega_3,\theta_2,\theta_3)-\omega_3.
\end{equation}
Obtaining the analytical expression for $\tilde{\omega}_2$ is difficult; therefore in all calculations we solve Eq.~(\ref{eq:pmcondition}) numerically or graphically. As a result,
\begin{multline}\label{eq:apsingles}
    \int_{-\infty}^{+\infty}\int_{-\infty}^{+\infty}\mathscr{N} \,d\vec{k}_{1}\,d\vec{k}_2= (2\pi)^4 R^{(3)} \frac{\omega_3 v_3}{c^2 n_3}\times \\ \int \frac{\tilde{\omega}_1(\tilde{\omega}_2,\omega_3,\theta_2,\theta_3) \tilde{\omega}_2^3(\omega_3,\theta_2,\theta_3)  \tilde{v}_1  \tilde{n}_2}{\tilde{n}_1^2 } \sin\theta_2 d\theta_2.
\end{multline}
To find the rate of photon emission within a certain detection bandwidth, $\Delta \omega_{det}$ in frequency and $\Delta \Omega_{det}$ in solid angle one should additionally integrate the differential  rate (\ref{eq:apsingles}) over these bandwidths.

\bibliographystyle{ieeetr}
\bibliography{bibliography}

\begin{thebibliography}{10}
\bibitem{note1}
``Dimensionality of $\gamma_s^{(n)}$ and $\gamma^{(n)}$ is not the same, as
  $\gamma_s^{(n)}$ accounts for the classical seed field amplitude, whereas
  $\gamma^{(n)}$ does not.''.
\end{thebibliography}

\end{document}